\documentclass[fleqn,12pt,twoside]{article}
\usepackage[headings]{espcrc1}

\readRCS
$Id: espcrc1.tex,v 1.2 2004/02/24 11:22:11 spepping Exp $
\usepackage{graphicx}

\usepackage[figuresright]{rotating}

\usepackage{bm}% bold math
\usepackage{epsfig}
\usepackage{color}

\newcommand{\AmS}{{\protect\the\textfont2
  A\kern-.1667em\lower.5ex\hbox{M}\kern-.125emS}}

\title{Particle production at forward rapidity in d+Au and Au+Au collisions in
STAR experiment at RHIC}

\author{Bedangadas Mohanty\address{Variable Energy Cyclotron Centre, 
        1/AF, Bidhan Nagar, \\ 
        Kolkata - 700064, India}%
        \thanks{For the full list of STAR authors and acknowledgments,
                      see appendix 'Collaborations' of this volume.}
        (for the STAR Collaboration) }
\runtitle{Particle production at forward rapidity}
\runauthor{B. Mohanty}

\begin{document}

% typeset front matter
\maketitle

\begin{abstract}
We present the recent results from the STAR experiment 
on charged and neutral particle measurements 
at the forward rapidity in d+Au collisions 
at $\sqrt{s_{\mathrm {NN}}}$ = 200 GeV and Au+Au collisions 
at $\sqrt{s_{\mathrm {NN}}}$ = 62.4 GeV. 
The nuclear modification factor for charged and neutral 
hadrons in d+Au collisions are presented. 
Measurements of $\Lambda$ and $\bar{\Lambda}$ production 
at forward rapidity and the variation of net baryon density 
as a function of collision centrality are discussed. We have 
also studied the limiting fragmentation of photons and charged 
particles in Au+Au collisions. The photons and charged particles 
separately follow the energy independent limiting fragmentation 
behaviour. However they have been observed to follow a different 
centrality dependence of limiting fragmentation.
\end{abstract}

\section{Introduction}
The STAR experiment at the Relativistic 
Heavy Ion Collider (RHIC) at the Brookhaven National 
Laboratory has a unique capability of precise 
measurement of charged and neutral hadrons and photon 
multiplicity at forward rapidity. Through this capability
we can carry out a systematic study of various aspects of particle 
production in relativistic heavy ion collisions. The study of 
nuclear modification factor ($R_{\mathrm {AB}}$) 
as a function of transverse momentum
($p_{\mathrm T}$) and pseudorapidity ($\eta$) can reveal 
interesting information on the mechanism of particle 
production. A $R_{\mathrm {AB}}$($p_{\mathrm T}$) of unity will indicate 
the absence of nuclear effects such as shadowing, multiple 
scattering (Cronin effect) and gluon saturation. In a color glass 
condensate picture (CGC) of particle production it is expected that the 
nuclear modification factor will decrease with the increase in 
rapidity~\cite{cgc}. Measurement of baryons at forward rapidity will help in 
understanding the baryon transport and quantifying the nuclear 
stopping in high energy collisions. Forward rapidity 
is also the region where we observe  the 
particle production per participating nucleon pair as a function 
of $\eta$ - y$_{\mathrm {beam}}$, where y$_{\mathrm {beam}}$ 
is the beam rapidity, to be independent of beam energy. 
This phenomenon is known as limiting fragmentation (LF)~\cite{starphoton}.
A detailed study of energy, centrality and 
species dependence of LF behaviour will provided useful insight to 
particle production mechanism in forward rapidity. Comparative study 
of LF for photons (primarily from decay of $\pi^{0}$ mesons) and 
charged particles will help in understanding the centrality dependence
LF behaviour.

\section{Detectors at forward rapidity in STAR}
The detectors at the forward rapidity~\cite{detectors} in STAR experiment are 
Forward Time Projection Chamber (FTPC), 
Photon Multiplicity Detector (PMD) and 
Forward $\pi^{0}$ Detector (FPD). 
The FTPCs detect charged particles in the pseudorapidity 
region 2.5 $<$ $\mid\eta\mid$ $<$ 4.0. The PMD consists of two 
planes of an array of cellular gas proportional counters 
separated by a lead convertor. It detects photons within 
2.3 $<$ $\eta$ $<$ 3.7. The FPD consists of lead-glass cells 21 
radiation lengths deep. It detects 
high energy $\pi^0$ mesons within $3.3< \eta < 4.1$. 

\section{Results from d+Au collisions at $\sqrt{s_{\mathrm {NN}}}$ = 200 GeV}
%----------------------------------------------------------------------
\begin{figure}
\begin{center}
\includegraphics[scale=0.37]{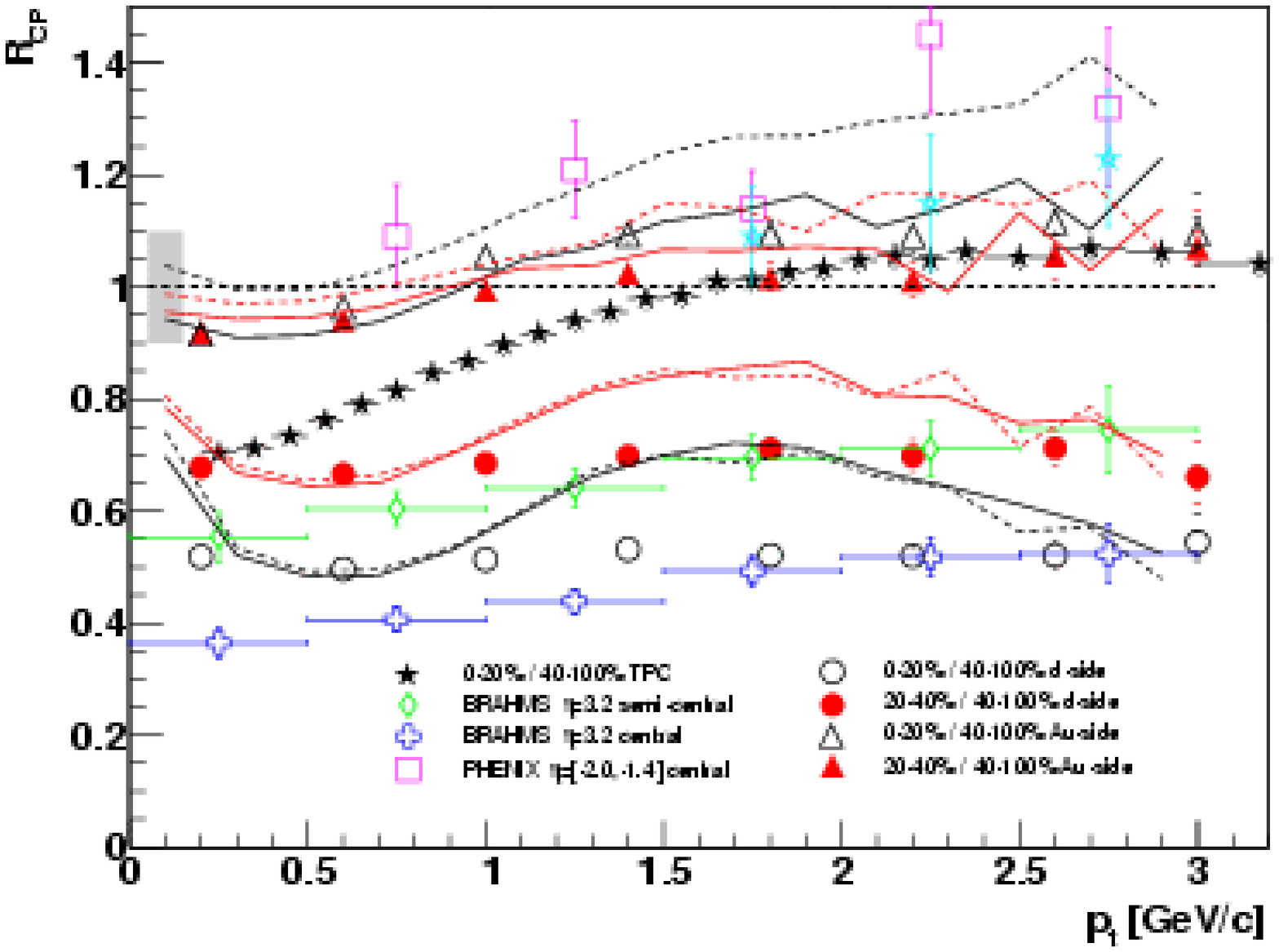}
\includegraphics[scale=0.28]{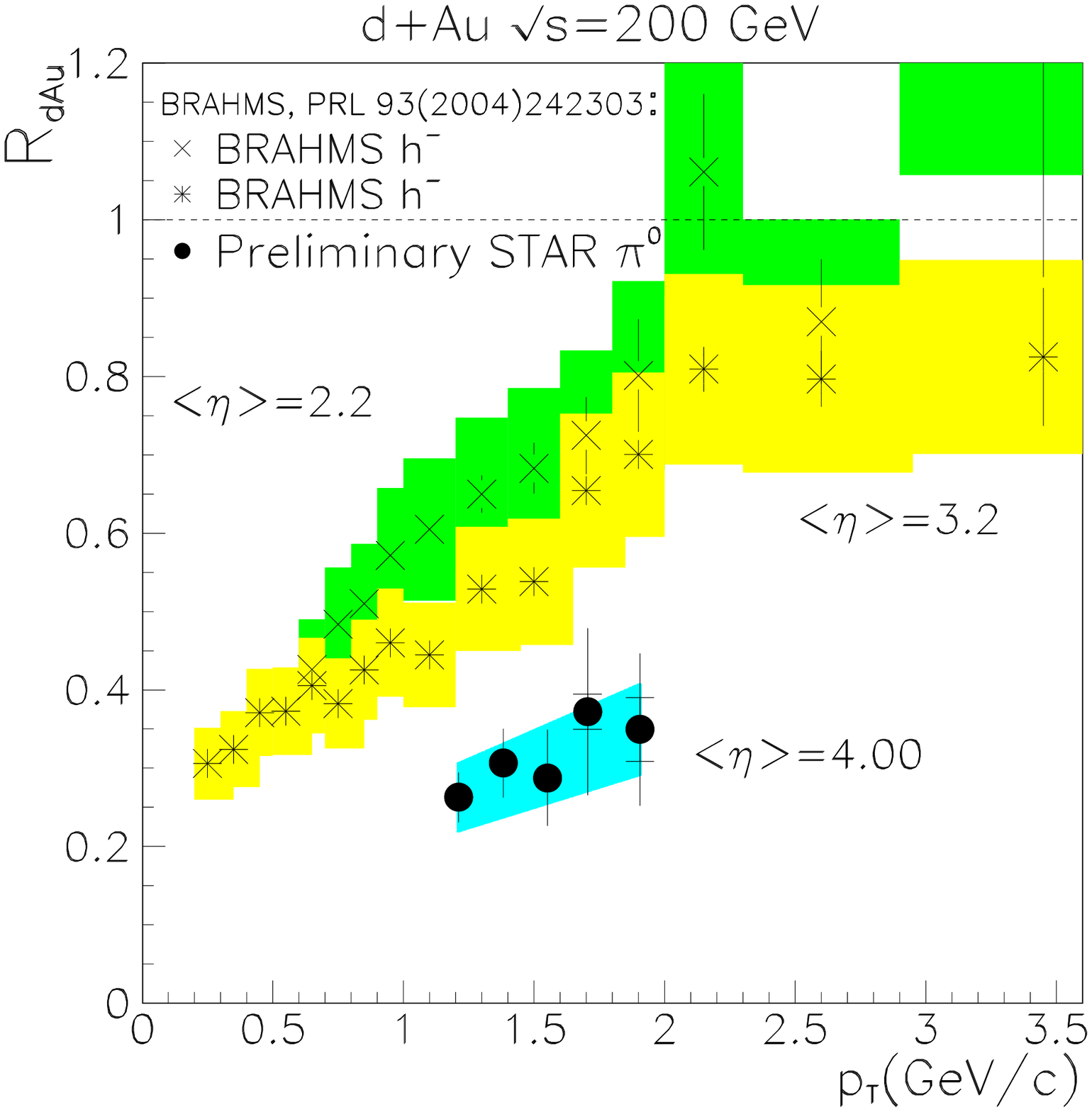}
\vskip -11.0mm
\caption{ Nuclear Modification factors for charged (left) and 
neutral (right) hadrons in d+Au collisions at $\sqrt{s_{\mathrm{NN}}}$ = 200 GeV.
The results are compared to BRAHMS and PHENIX results.
The charged hadron measurements are also compared to HIJING. 
}
\label{nmf}
\end{center}
\end{figure}
%----------------------------------------------------------------------
\subsection{Charged and neutral hadron production}
In Fig.~\ref{nmf} (left) we show the nuclear modification factor 
($R_{\mathrm {CP}} = \frac{(d^{2}N/dp_{t}d\eta/<N_{\mathrm {bin}}>)\mid_{\mathrm {central}}}
{(d^{2}N/dp_{t}d\eta/<N_{bin}>)\mid_{\mathrm {peripheral}}}$) 
of charged hadrons for two centrality classes 
at $\mid \eta \mid$ $\sim$ 3.1 and up to $p_{\mathrm T}$ = 3 GeV/$c$.
$R_{\mathrm {CP}}$ is larger on the gold side than on 
the deuteron side. 
%This indicates that the Cronin effect is more 
%the on gold side. 
On the deuteron side we observe a stronger centrality 
dependence of $R_{\mathrm {CP}}$ compared to the gold side. Our measurement 
compares well with the results from BRAHMS and PHENIX 
experiments~\cite{brahms_phenix}. The lines are calculations
from HIJING with (dashed) and without (solid) shadowing.
The nuclear effects on particle production have been also quantified 
by studying the $R^{\pi^{0}}_{\rm {dAu}}$ (Fig.~\ref{nmf} (right)), the ratio of 
the inclusive yield of $\pi^{0}$ in d+Au to p+p collisions normalized by the 
number of nucleon-nucleon collisions. The ratio $R^{\,\pi^0}_{\rm {dAu}}$ 
at $\langle\eta\rangle =4.0$ is significantly smaller 
than $R^{\,h^-}_{\rm {dAu}}$ at smaller $\eta$~\cite{brahms_phenix},
consistent with the trend expected from CGC-based models~\cite{cgc}.

%----------------------------------------------------------------------
\begin{figure}
\begin{center}
\includegraphics[scale=0.37]{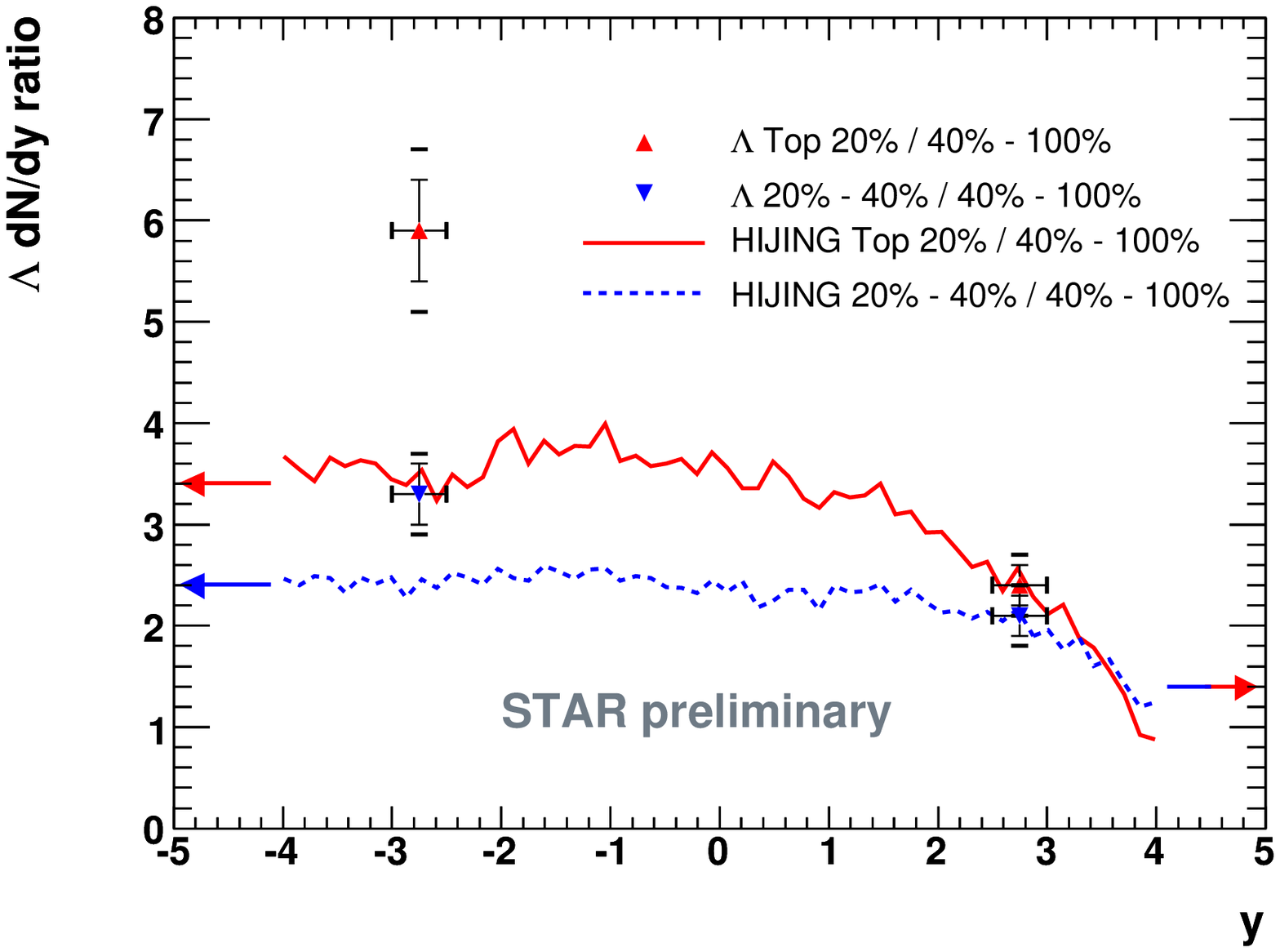}
\includegraphics[scale=0.37]{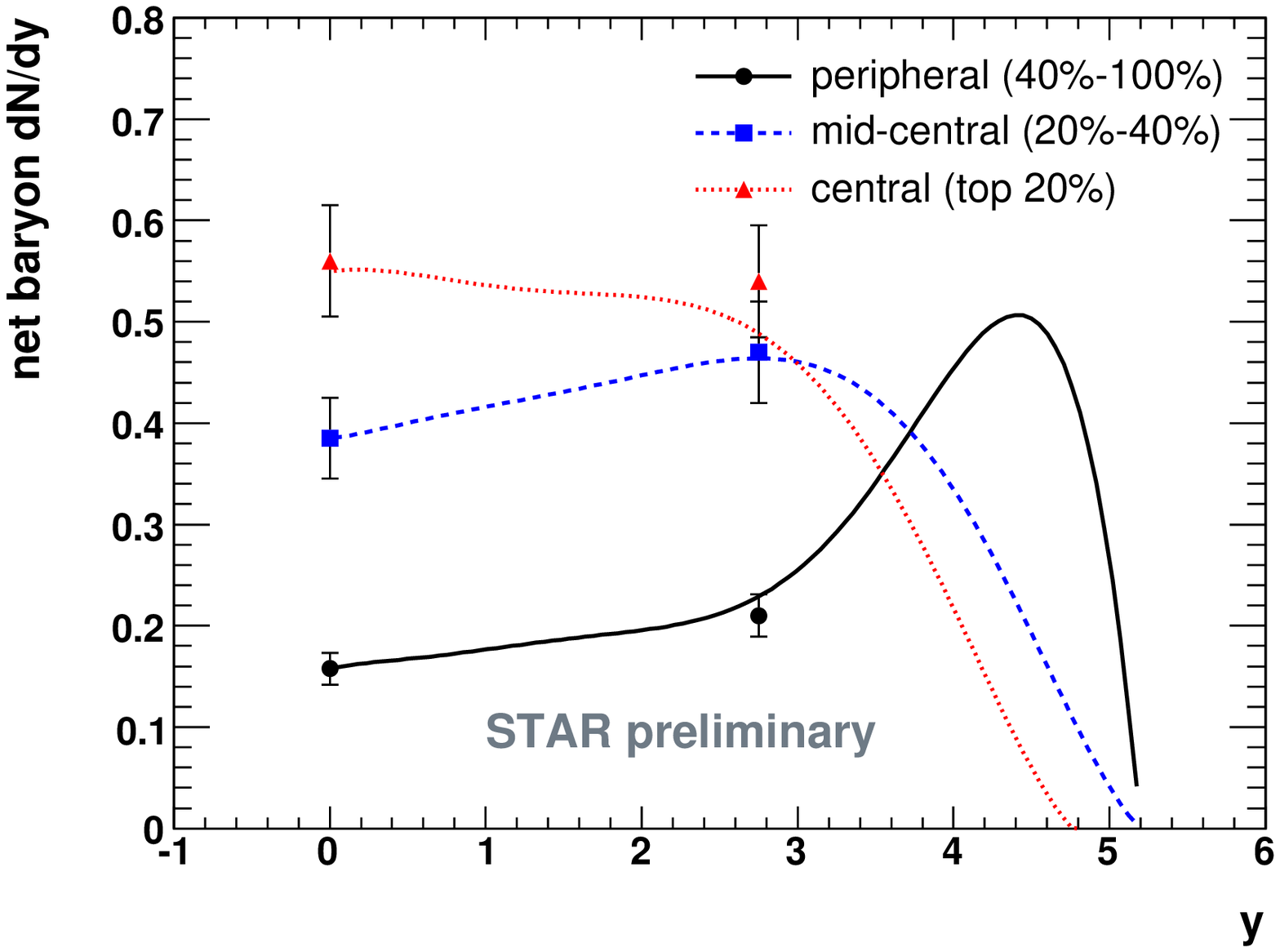}
\vskip -13.0mm
\caption{Left: Ratio of the $\Lambda$ yields in central and peripheral 
collisions compared to HIJING. Right: Net baryon density for three centrality
classes on the deuteron side of the collision.
}
\label{lambda}
\end{center}
\end{figure}
%----------------------------------------------------------------------
\subsection{Strange particle production and baryon stopping}
In the FTPC's, $\Lambda$ and $\bar{\Lambda}$ are reconstructed 
by obtaining the invariant mass of their dominant decay modes
$\Lambda \rightarrow p \pi^{-}$ and $\Lambda \rightarrow \bar{p} \pi^{+}$
(branching ratio of 64\%). In Fig.~\ref{lambda} (left) we show the ratio of
the $\Lambda$ rapidity density in central collisions to that in 
peripheral collisions
as a function of rapidity. The data show significant deviations 
from expectations from wounded nucleon scaling (arrows).
The results when compared to HIJING calculations show a good agreement 
on the deuteron side of the collisions. On the gold side the data is 
above HIJING indicating possible presence of substantial nuclear effects. 
In Fig.~\ref{lambda} (right)
we show the measured net baryon density for three centrality classes. 
The midrapidity results are from measured p and $\bar{p}$ in the 
Time-Of-Flight detector and at y = 2.75 on the deuteron side from 
the $\Lambda$ measurements in FTPC. 
Lines show fits to the net baryon distributions with global
constraints on the total number of baryons.
%The fits represent the possible
%net baryon distribution as a function of rapidity. 
Within such a model, we observe 
significant stopping in central collisions and a large degree of
transparency in peripheral collisions.

%----------------------------------------------------------------------
\begin{figure}
\begin{center}
\includegraphics[scale=0.38]{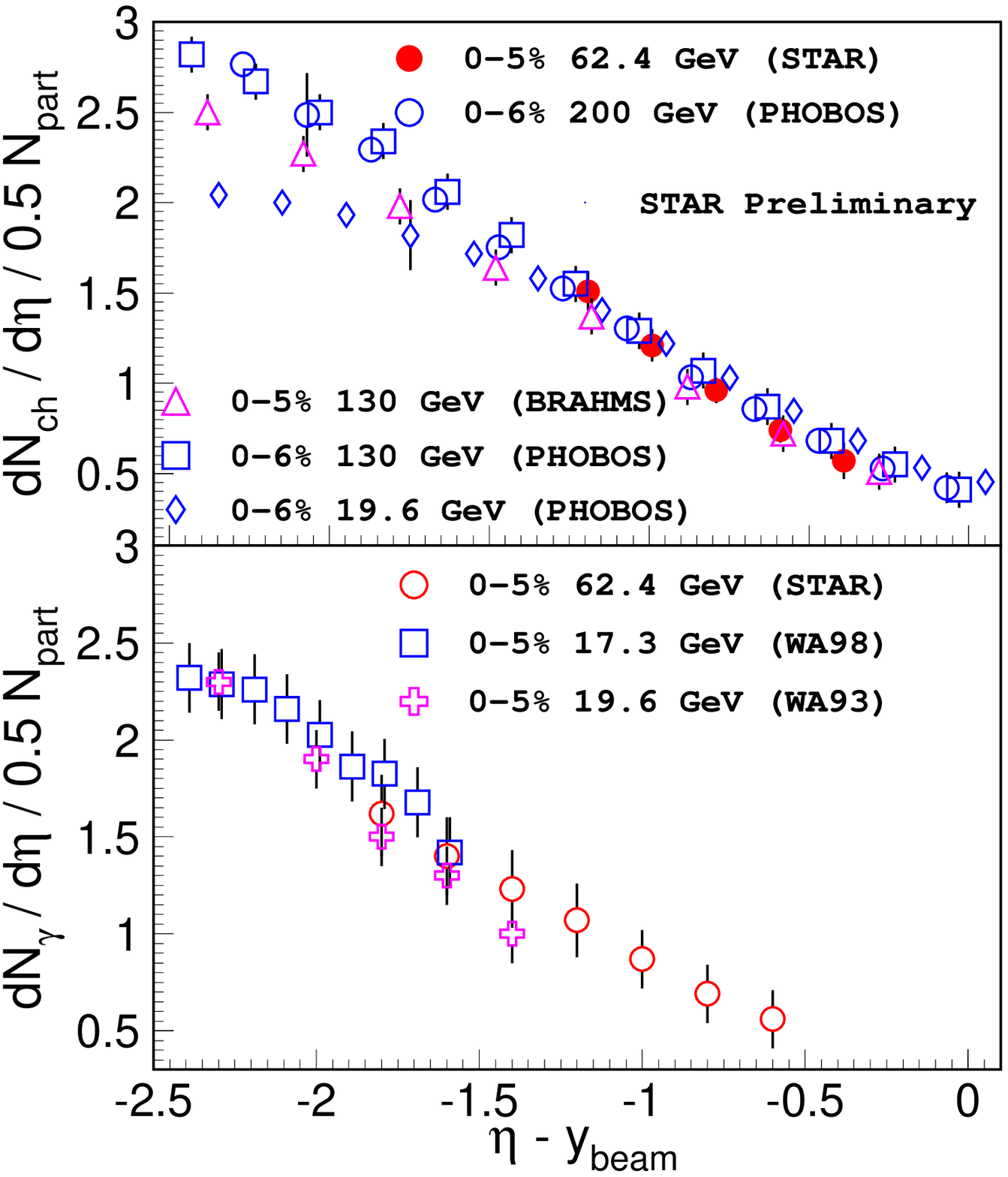}
\includegraphics[scale=0.38]{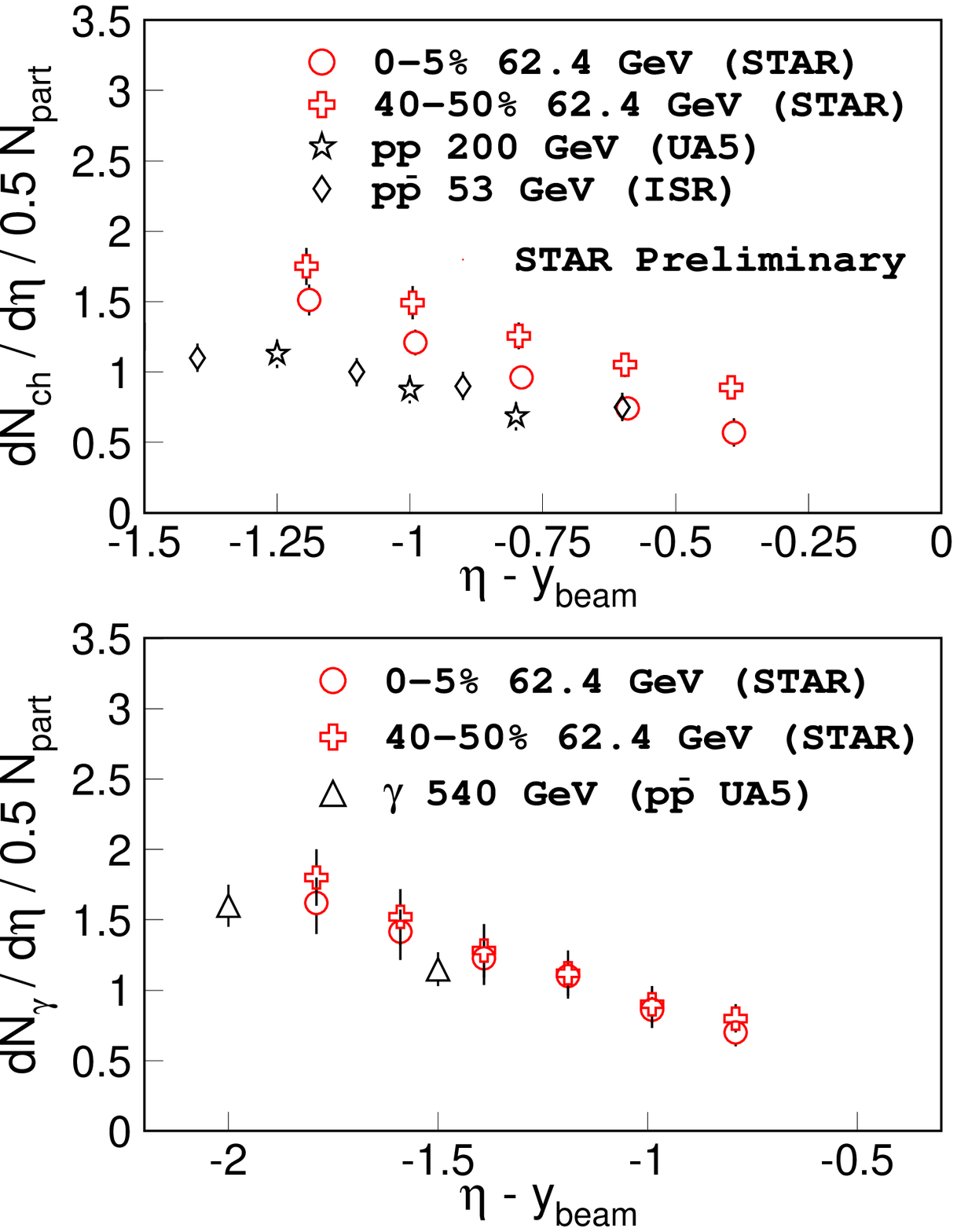}
\vskip -14.0mm
\caption{ $\frac{d N}{d \eta}$ for charged particles 
and photons~\cite{starphoton} 
for various collision systems : (left)  at different 
$\sqrt{s_{\mathrm {NN}}}$ and (right) for various event centrality classes 
at $\sqrt{s_{\mathrm {NN}}}$  = 62.4 GeV.
}
\label{lim}
\end{center}
\end{figure}
%----------------------------------------------------------------------
\section{Results from Au+Au collisions at $\sqrt{s_{\mathrm {NN}}}$ = 62.4 GeV}
In Fig.~\ref{lim} we present 
the energy dependence of LF for inclusive photons and 
charged particles. Our data from central collisions has been compared to 
corresponding measurements from nucleus-nucleus 
collisions at different $\sqrt{s_{\mathrm {NN}}}$ 
in PHOBOS, BRAHMS, WA98, WA93 experiments and p+p($\bar{p}$)
collisions in ISR and UA5 experiments~\cite{starphoton}. 
The  $\frac{d N}{d \eta}$ per participant 
pair as a function of $\eta$ - y$_{\mathrm {beam}}$ is observed to be 
independent of beam energy.
In Fig.~\ref{lim} we show the centrality dependence of LF 
for charged particles and photons.
At forward rapidity in Au+Au collisions at $\sqrt{s_{\mathrm {NN}}}$ 
= 62.4 GeV the charged particle yield normalized to number of 
participating nucleons ($N_{\mathrm {part}}$) 
as a function of $\eta$ - y$_{\mathrm {beam}}$ is 
higher for peripheral collisions compared to central collisions, 
whereas within the measured $\eta$ range of 2.3 to 3.7, the photon yield 
normalized to $N_{\mathrm {part}}$ as a function 
of $\eta$ - y$_{\mathrm {beam}}$ is independent of centrality~\cite{starphoton}.
We also observe that the photon results in the forward rapidity 
region from $p \bar{p}$ collisions at $\sqrt{s_{\mathrm {NN}}}$ = 540 GeV 
are in close agreement with the measured photon yield in Au+Au collisions 
at $\sqrt{s_{\mathrm {NN}}}$ = 62.4 GeV. However, the $pp$ and $p \bar{p}$ inclusive 
charged particle results are very different from those for 
Au+Au collisions at $\sqrt{s_{\mathrm {NN}}}$ = 62.4 GeV. 
The photon results and comparison with p+p($\bar{p}$) collisions 
indicate that, in the $\eta$ region studied, there is apparently a 
significant charged baryon contribution in Au+Au collisions 
and such a contribution may be 
responsible for the observed centrality dependent 
LF behaviour in charged particles.

\section{Summary}
We have presented the results from the STAR experiment on charged hadrons, 
neutral hadrons and photons at forward rapidity in 
d+Au collisions at $\sqrt{s_{\mathrm {NN}}}$ = 200 GeV and Au+Au collisions 
at $\sqrt{s_{\mathrm {NN}}}$ = 62.4 GeV. The nuclear modification factor 
in d+Au collisions is observed to decrease with increasing 
rapidity, which is consistent with predictions from CGC based models. 
$R_{\mathrm {CP}}$ is larger on the gold side than on 
the deuteron side in d+Au collisions. 
From the net baryon density
measurements we conclude that there is significant stopping in central 
d+Au collisions  and large transparency in peripheral collisions.
Our measurements of photons and charged particle multiplicity show
that photons follow centrality independent limiting fragmentation
behaviour whereas charged particles follow a centrality dependent behaviour.

\end{document}